\documentstyle[prl,aps,psfig]{revtex}

\def\sqr#1#2{{\vcenter{\hrule height.#2pt\hbox{\vrule width.#2pt
height#1pt \kern#1pt \vrule width.#2pt}\hrule height.#2pt}}}

\newcommand{\be}{\begin{equation}}
\newcommand{\ee}{\end{equation}}
\newcommand{\ben}{\begin{eqnarray}}
\newcommand{\een}{\end{eqnarray}}
\newcommand{\bc}{\begin{center}}
\newcommand{\ec}{\end{center}}

\begin{document}
\preprint{Sussex preprint SUSSEX-AST 97/4-1, gr-qc/9704029}
\draft
\twocolumn[\hsize\textwidth\columnwidth\hsize\csname
@twocolumnfalse\endcsname

\title{\bf The gravitational redshift of boson stars}

\author{Franz E.~Schunck and Andrew R.~Liddle}

\address{Astronomy Centre, University of Sussex, Falmer, Brighton BN1 9QJ,
United Kingdom\\
{\tt fs@astr.maps.susx.ac.uk , a.liddle@sussex.ac.uk}}

\date{\today }

\maketitle

\begin{abstract}
We investigate the possible gravitational redshift values for boson
stars with a self-interaction, studying a wide range of possible
masses. We find a limiting value of $z_{\rm lim} \simeq 0.687$ for
stable boson star configurations. We compare theoretical expectation
with the observational capabilities in several different wavebands,
concluding that direct observation of boson stars by this means will
be extremely challenging. X-ray spectroscopy is perhaps the most
interesting possibility.
\end{abstract}

\bigskip

\pacs{PACS no.: 04.40.Dg, 98.54.-h \hspace*{1.5cm} Sussex preprint
SUSSEX-AST 97/4-1, gr-qc/9704029}

\vskip2pc]


\section{Introduction}

The idea of the boson star goes back to Kaup in 1968 \cite{kau}. A
boson star is a gravitationally bound collection of bosonic particles,
arising as a solution of the Klein--Gordon equation coupled to general
relativity. Many investigations of the possible configurations have
been carried out; for reviews see Ref.~\cite{rev}. For
non-self-interacting bosons of mass $m$, the mass of a typical
configuration is of order $m^2_{{\rm Pl}}/m$, to be compared with a
typical neutron star mass of $m^3_{{\rm Pl}}/m^2_{{\rm neutron}}$,
which is about a solar mass. Here $m_{{\rm Pl}}$ is the Planck
mass. Unless the bosons are extremely light, boson stars made from
non-interacting bosons are orders of magnitude lighter than typical
stars.

The situation is very different if the boson stars have even a very
weak self-interaction; for a first investigation of self-interactions
see Ref.~\cite{mie}. Colpi, Shapiro and Wasserman \cite{col} showed
that the maximum mass of stable configurations is then of order
$\lambda^{1/2} m^3_{{\rm Pl}}/m^2$, where $\lambda$ is the scalar
field self-coupling, normally assumed to be of order unity. Then boson
star configurations exist with mass (and radius) similar to that of
neutron stars, if the bosons, like neutrons, have a mass around 1
GeV. They can also be much heavier, should the bosons be lighter. In
this paper, we allow ourselves to consider a very wide range of
possibilities for the boson star mass and radius.

It remains unknown whether or not boson stars could actually form in
our own Universe, for example through a gravitational instability
process, though there has been some discussion in the literature
\cite{form}. The most favourable situation would be for the dark
matter in the Universe to be in the form of very weakly interacting
bosons, a substantial fraction of which manage to form boson stars;
such a process may be aided if the bosons have a reasonably strong
coupling to each other, but not to conventional matter. If boson stars
can form, they provide an alternative explanation for stellar systems
in which an object is inferred to have a high mass; conventionally, a
`star' with mass greater than a few solar masses is assumed to be a
black hole (since this is above the maximum permitted mass for a
neutron star), but a boson star offers a more speculative alternative.

Dark matter is characterized by its extremely weak interaction with
conventional matter; even though 90\% of the mass of the galaxy is in
this form, it is not directly visible. In this paper, we investigate
the implications of assuming that the material from which boson stars
are made is similarly extremely weakly interacting, so that its only
interaction with neighbouring baryonic material and photons is
gravitational, just as the relation between the visible galaxy and its
dark matter halo.\footnote{An example already existing in the
literature is the {\em boson--fermion star}
\cite{bosfer}, which is made up of bosons and neutrons interacting only 
gravitationally.} However, while a galaxy halo can be described using 
Newtonian theory, boson stars close to the 
maximum allowed mass are general relativistic objects. This gives a new 
characteristic of such objects, a gravitational redshift.
In this paper, we examine whether radiating baryonic matter moving in the 
gravitational potential of a boson star could be used as an observational 
signal of boson stars.

\section{The Einstein-Klein-Gordon equations}

The Lagrange density of a massive complex self-gravitating scalar
field reads
\be
{\cal L} = \frac {1}{2} \sqrt{\mid g \mid} \left[
  \frac {m_{{\rm Pl}}^2}{8\pi} R + \partial_\mu \Phi^\ast \partial^\mu \Phi 
- U(|\Phi 
|^2) \right ] \; , \label{lagr}
\ee
where $R$ is the curvature scalar, $g$ the determinant of the metric
$g_{\mu \nu }$, and $\Phi$ is a {\em complex} scalar field with a
potential $U$. We take $\hbar=c=1$. Then we find the coupled system of
equations
\ben
R_{\mu \nu } - \frac{1}{2} g_{\mu \nu } R & = &
                  - \frac{8\pi}{m_{{\rm Pl}}^2} T_{\mu \nu } (\Phi ) \; , \\
\Box \Phi + \frac{dU}{d|\Phi |^2} \Phi & = & 0 \; ,
\een
where
\begin{eqnarray}
T_{\mu \nu } & = & (\partial_\mu \Phi^\ast ) (\partial_\nu \Phi )
	\nonumber \\
& &  - \frac{1}{2} g_{\mu \nu }
 \Bigl [ g^{\sigma \rho } (\partial_\sigma \Phi^\ast )
         (\partial_\rho \Phi ) - U(|\Phi |^2) \Bigr ]
\end{eqnarray}
is the energy-momentum tensor and
\be
\Box = \partial_\mu
\Bigl [ \sqrt{\mid g \mid } \; g^{\mu \nu } \partial_\nu \Bigr ]/ 
\sqrt{\mid g \mid }
\ee
the generally covariant d'Alembertian. Within the model we want to
have an additional global $U(1)$ symmetry, so we can take the
following potential
\be
U = m^2 |\Phi |^2 + \frac{\lambda }{2} |\Phi |^4 \; ,
\ee
where $m$ is the scalar mass and $\lambda$ a dimensionless constant
measuring the self-interaction strength.

The method for finding solutions is well known \cite{kau,rev}. For
spherically symmetric solutions we use the static line element
\be
ds^2 = e^{\nu (r)} dt^2 - e^{\mu (r)} dr^2
  - r^2 ( d\vartheta^2 + \sin^2\vartheta \, d\varphi^2) \; .
\ee
The most general scalar field ansatz consistent with this metric is
\be
\Phi (r,t) = P(r) e^{-i \omega t} \; ,
\ee
where $\omega$ is the frequency. Notice that to find finite mass
solutions, the scalar field must carry a time-dependence of this form,
which leaves the energy-momentum tensor, and hence the metric,
time-independent.

The non-vanishing components of the energy-momentum tensor are
\ben
T_0{}^0 = \rho &=& \frac{1}{2} [ \omega^2  P^2(r) e^{-\nu }
   + P'^2(r) e^{-\mu } + U ] \; , \\
T_1{}^1 = p_r &=& \frac{1}{2} [ \omega^2  P^2(r) e^{-\nu }
   + P'^2(r) e^{-\mu } - U ] \; , \\
T_2{}^2 = T_3{}^3 & = & p_\bot \quad \mbox{with} \nonumber \\
p_\bot & = &  - \frac{1}{2} [ \omega^2  P^2(r) e^{-\nu }
   - P'^2(r) e^{-\mu } - U ]
\een
where $'=d/dr$. Note that the pressure is anisotropic; there are two 
equations of state
$p_{r} = \rho - U$ and $p_\bot = \rho - U - P'^2(r) e^{-\mu }$.

The decisive non-vanishing components of the Einstein equation are
\ben
\nu' + \mu' & = & \frac{8\pi}{m_{{\rm Pl}}^2} (\rho + p_r) r e^\mu  
\; , \label{nula}\\
\mu' & = &  \frac{8\pi}{m_{{\rm Pl}}^2} \rho r e^\mu - \frac {1}{r} 
(e^\mu - 1)
\; , \label{la}
\een
the two further components giving degenerate equations because of
the Bianchi identities.

The differential equation for the scalar field is
\be
P'' + \left ( \frac {\nu' - \mu'}{2} + \frac {2}{r} \right )
 P' + e^{\mu - \nu } \omega^2 P
- e^{\mu } \frac{dU}{dP^2} P = 0  \; . \label{2ska}
\ee

For the rest of our paper, we employ the dimensionless quantities $x=m
r$, $\sigma = \sqrt{4\pi} \; P/m_{{\rm Pl}}$ and $\Lambda = \lambda
m_{{\rm Pl}}^2/4\pi m^2$. In order to obtain solutions which are
regular at the origin, we must impose the boundary conditions $\sigma
'(0)=0$ and $\mu (0)=0$.

Once the fundamental parameter $\Lambda$ has been fixed, the only
remaining freedom is the value of the scalar field at the centre of
the star (the scalar field mass simply contributing an overall
scaling); the initial value $\nu (0)$ is determined by the boundary
condition of Minkowskian spacetime at infinity. It is well known that
as the central scalar field value is increased, initially the inferred
mass of the star (determined from the asymptotic form of the metric)
increases up to a maximum, after which it exhibits oscillatory
behaviour, shown in Figure \ref{fig1}.  The configurations up to this
maximum mass are dynamically stable, but those beyond it are unstable
\cite{stab,kus,sch,rev} and should not be considered physical.

\section{Gravitational redshift}

In this Section we calculate the redshift produced by the
gravitational potential of a boson star. Under our assumption that the
scalar particles have no interaction with the baryonic matter other
than gravitationally, the baryonic matter can penetrate up to the
center of the boson star. If the baryonic matter emits or absorbs
radiation within the gravitational potential, the spectral feature
will be redshifted.

The gravitational redshift $z$ within a static line element is given by
(see e.g.~Ref.~\cite{nar})
\be
1+z_{{\rm g}} = \sqrt{\frac{e^{\nu (R_{{\rm ext}})}}{e^{\nu (R_{{\rm 
int}})}}} \; ,
\ee
where the emitter and receiver are located at $R_{{\rm int}}$ and $R_{{\rm
ext}}$ respectively.  The maximum possible redshift for a given
configuration is obtained if the emitter is exactly at the center $R_{{\rm
int}}=0$. The receiver is always practically at infinity, hence $\exp \left[
\nu (R_{{\rm ext}})
\right]=1$. For all 
other redshifts in between, we define the redshift function
\be
1+z_{{\rm g}}(x) \equiv \exp \left(-\frac{\nu (x)}{2}\right)  \,.
\label{zfunc} 
\ee
Table \ref{table1} gives the results for the maximum mass boson star
for different $\Lambda $. The maximum mass gives the highest value one
can obtain from stable configurations (unstable configurations can
yield very high redshift values --- we show one as an example).  One
recognizes that with increasing self-interaction values $\Lambda $
also the maximal redshift value grows. For unstable states very high
$z_g$ values can be found.  For $\Lambda \rightarrow \infty $, we find
the asymptotic value $z_{{\rm lim}} \simeq 0.687$, as shown in
Fig.~\ref{fig2}.

Figure \ref{fig3} shows the decrease of the asymptotically measured
redshift as a function of radius, for different $\Lambda $ values.

The highest central value is quite interesting. In the early years of
quasar detection, it was speculated whether the origin of the redshift
could be gravitational rather than cosmological; see, for example,
Refs.~\cite{bur,hoy,gre}. The Schwarzschild interior solution for a
perfect fluid sphere with constant density gives a maximal redshift on
the surface of 2 (cf.~\cite{buch}), but this matter form is
unrealistic because it has a transmission velocity exceeding $c$.  The
value 2 is also an upper limit for all static fluid spheres whose
density does not increase outwards \cite{buch}.  Bondi \cite{bon}
applied constraints on a perfect fluid with isotropic pressure and
looked for stable objects; from these results, a maximal surface
redshift of $z_{\mbox{max}} \simeq 0.62$ can be derived, a value which
is quite close to our central one of 0.687 for the anisotropic boson
matter fluid.  For a neutron star, one finds a central value of 0.47
\cite{mar}, whereas for a star consisting of two kinds of fermions one
has a maximal central redshift (for a stable configuration) of 1.22
\cite{zha}.  Different forms of matter result in completely
different central redshift values. Just a small self-interaction
$\Lambda =1$ for the bosonic particles gives a redshift value about
0.49, just above that of the neutron star.

In general, observed redshift values could consist of a combination of
cosmological and gravitational redshift. Their relation is given by
\be
1+z = (1+z_{{\rm c}}) (1+z_{{\rm g}}) \; ,
\ee
because of the relativistic velocity addition formula. The objects
with highest redshift values today are the quasars with redshifts up
to about five. If one assumes that they would have an incorporated
gravitational redshift, and our solutions give a maximal value of
about 0.7, then we have a cosmological redshift of 2.53.  Even if for
some reason they possessed a maximal gravitational redshift, this
could not entirely explain the observed redshifts. In any case, there
are no indications of a non-cosmological redshift for quasars at
present.

\section{Detectability of boson stars}

The mass $M$ of a boson star composed of non-self-interacting
particles is inversely proportional to $m$, while the mass of a
self-interacting boson star is proportional to $\sqrt{\lambda }/m^2$
\cite{col}. Taking $\lambda \sim 1$, then for small $m$ (to be
precise, provided $\Lambda \gg 1$) the slef-interacting star is much
more massive.  For example, if we want to get a boson star with a mass
of order $10^{33}$g (a solar mass), then we need $m \sim 10^{-10}$ eV
for $\lambda = 0$, or $m \propto \lambda^{1/4}$ GeV if $\lambda \gg
10^{-38}$ (we see that the self-coupling has to be extraordinarily
tiny to be negligible). In this example, the scalar particle has a
mass comparable to a neutron, leading to a boson star with the
dimensions of a neutron star. If we reduce the scalar mass further, to
$m \sim 1$ MeV, then we find $M \sim 10^{39} \sqrt{\lambda }$ g and
$R\sim 10^6 \sqrt{\lambda }$ km; this radius is comparable to that of
the sun, but encloses 10$^6$ solar masses. These parameters are
reminiscent of supermassive black holes, for example as in Active
Galactic Nuclei; the mass--radius relation is effectively fixed just
by the objects being relativistic. An exceptionally extreme example is
to take $m \sim 1$ eV, giving $M \sim 10^{18} \sqrt{\lambda } M_\odot$
and $R\sim 10^3 \sqrt{\lambda }$ parsecs.

These numbers must be compared with the critical matter density in the
Universe, around $10^{12} M_\odot \, {\rm Mpc}^{-3}$. This sets the
scale of the typical distance expected to the nearest object of a
given mass, provided such objects are common. The largest
gravitationally bound objects in the present Universe are rich galaxy
clusters, with masses of order $10^{15} M_\odot$, which require the
assembly of material from a volume around $10$ Mpc on a side.

In all cases, the density of the boson stars makes their direct
detection difficult; in particular, they cannot be resolved in any
waveband. The best optical spectroscopy has a resolution of order of
an arcsecond, nowhere near good enough. The best imaging resolution is
in the radio band; for example the VSOP program \cite{halca}, which
includes the recently launched Japanese HALCA satellite, offers
resolution of around a milli-arcsecond, but this approach is limited
because spectroscopy cannot be carried out. At a distance of 10Mpc, we
could resolve a boson star with a radius of 0.1pc, corresponding to a
mass of $10^{15}M_\odot$, but rich clusters cannot possibly be
dominated by a single massive object of that sort.  Any reduction of
the boson star mass requires it to be closer, in proportion to the
mass $M$, but the amount of mass available in that volume is
decreasing as the cube of the distance. The situation therefore
rapidly becomes worse at closer distances; the nearest object of a
given size is expected to be far too distant to be resolvable.  For
example, to be resolvable at a distance of one kiloparsec, a boson
star would need a radius of 10$^{13}$cm (roughly the Earth--Sun
separation) requiring a mass of $10^{11} M_\odot$, almost the mass of
an entire galaxy.

However, even if boson stars cannot be directly resolved, their
influence might still be visible if material in their vicinity is
sufficiently luminous. It is necessary to find a certain amount of
luminous matter within the gravitational potential of the boson star.
This could, for example, be H{\small I} gas clouds as seen in
galaxies.  {}From the rotation curves calculated below, one imagines
that the hydrogen gas would probably be excited, hence H{\small II},
because of the kinetic energy gain in the gravitational potential of
the boson star.  One might also expect accretion discs about boson
stars, though there the luminosity could be dominated by regions
outside the gravitational potential and the boson star would be
indistinguishable from any other type of compact object.  One should
mention that bright H{\small II} regions are found within the host
galaxies of quasars \cite{bah}.

A more dramatic possibility would be to house a complete nuclear
burning star within the gravitational potential of the boson
star. Unfortunately that does not seem feasible, since we have already
seen that to have a solar radius the boson star must weigh $10^6
M_\odot$. A more generic situation therefore would be for lighter
boson stars to be contained completely within conventional stars,
influencing the stellar structure by means of their additional
gravitational interaction. Such a situation has in fact been studied
for a degenerate neutron core inside a supergiant, the so-called
Thorne--\.{Z}ytkow objects \cite{TZ}. If the bosons are completely
non-interacting, then the boson star might indeed be able to survive
such an environment. With an interaction present, one would have to
examine whether or not the boson star could be dissociated by high
energy photons in the stellar interior, so that instead the scalar
particles circulate freely within the star. Historically, such a
situation was proposed as a possible resolution of the solar neutrino
problem, under the name {\em cosmion}
\cite{cosmion}.

The results from the last section showed that a boson star, like a
conventional star, cannot be resolved in itself but rather would have
to be part of a larger area. We know that the luminosity from the
boson star will be redshifted. Hence we can distinguish between the
luminosity of the boson star and other areas if there is a
particularly strong spectral line.  The luminosity $L_{{\rm boson}}$
could be derived from the redshifted tail of the emission line
subtracting the `normal' part.  First, we assume that the gas
distribution is everywhere homogeneous within the resolved area. This
could be, in first order, valid for almost Newtonian boson stars. Let
$L$ be the complete measured luminosity; then we have $L=L_{{\rm
normal}}+L_{{\rm boson}}$, where we divided $L$ into its two parts of
the `normal' non-redshifted luminosity and the one from the boson star
area. Both parts are proportional to the product of the relation $A$
of the boson star area to the total resolved area and the total
luminosity $L$, hence, $L_{{\rm boson}}=A L$ while $L_{{\rm
normal}}=(1-A) L$. The calculated value of $A$ gives only an upper
limit for the size of the boson star.

Given the inability to resolve boson stars, the most promising
technique is to consider a wave-band where they might be extremely
luminous. The best example is in X-rays. A very massive boson star,
say $10^6 M_{\odot}$ is likely to form an accretion disk and since its
exterior solution is Schwarzschild it is likely to look very similar
to an AGN with a black hole at the center. In X-rays it may be
possible to probe very close to the Schwarzschild radius; it has been
claimed by Iwasawa et al.~\cite{Ietal} that using ASCA data they have
probed to within 1.5 Schwarzschild radii. Because this is inside the
static limit for a non-rotating black hole, they conclude that a Kerr
geometry is required. We note that a boson star configuration provides
a more speculative alternative, giving a non-singular solution where
emission can occur from arbitrarily close to the center.  The
signature they use is a redshifted wing of the Iron K-line, indicative
of emission from deep within the gravitational potential well. If such
techniques have their validity confirmed, it may ultimately be
possible to use X-ray spectroscopy to map out the shape of the
gravitational potential close to the Schwarzschild radius or boson
star, and perhaps to differentiate between the theoretical predictions
for a Kerr black hole and for a rotating boson star.

\section{Rotation curves}

We end by considering the rotation curves about a boson star in more
detail.  For the static spherically symmetric metric considered here
circular orbit geodesics obey
\ben
v_\varphi^2 & = & r \nu' e^{\nu }/2 = e^{\nu } (e^{\mu}-1)/2 +
	\frac{8\pi}{m_{{\rm Pl}}^2} p_r r^2 e^{\mu +\nu }/2 
	\nonumber \\
 & \simeq &
	M(r)/r + \frac{8\pi}{m_{{\rm Pl}}^2} p_r r^2 e^{\mu +\nu }/2
	\; ,
\een
which reduces for a weak gravitational field into the Newtonian form
$v_{\varphi,{\rm Newt}}^2 = M(r)/r$ if $p_r=0$.

Figure \ref{fig4} shows the rotation curves for the cases $\Lambda =0$
and 300.  The curves increase up to a maximum followed by a Keplerian
decrease. The possible rotation velocities circulating within this
gravitational potential are quite remarkable.  Their maximum reaches
more then one-third of the velocity of light; if boson stars can form,
then such enormous rotation velocities are not necessarily signatures
of black holes. The matter possesses an impressive kinetic energy of
about 6\% of the rest mass.  If we were to suppose that each year a
mass of 1$M_\odot $ transfers this amount of kinetic energy into
radiation, a boson star had a luminosity of 10$^{44}$erg/s.

Figure \ref{fig5} shows how the rotation velocities decrease with
decreasing initial value $\sigma(0)$ at fixed $\Lambda$.  Table
\ref{table2} gives the maximal velocity. For Newtonian solutions the
rotation velocity is low and quite constant over a larger interval.

\section{Discussion}

In studying a wide range of possible masses and coupling for the
bosons, we have unveiled a range of possibilities for boson stars to
have an observational impact. In the literature it is common to
consider boson stars to be like neutron stars, but this relies on
specific choices for the boson mass for which there is no strong
motivation.  A boson star could have quite different dimensions both
in size and mass.  In analogy to the dark matter halos of galaxies, a
boson star could also be {\em transparent}, i.e.~without any
electro-weak and strong interactions.  Because of the general
relativistic background of the boson star solutions, a gravitational
redshift has to be taken into account. We find that for stable boson
stars a certain redshift value cannot be exceeded.  This is in good
agreement with results for general fluids and especially the neutron
star model where one finds similar redshift values. We mention that if
one considers instead bosonic particles without mass or
self-interaction, one finds solutions which can be applied to fit
rotation curve data of spiral and dwarf galaxies, and which may also
give very high redshift values \cite{sch2}.

It will be observationally very challenging to use the gravitational
potential of these configurations to provide an observational
test. Although radio offers by far the best imaging resolution, it
seems that it may be spectroscopy using X-rays which offers the best
opportunity to probe the central regions of strong gravitational
sources.

\acknowledgments
FES is supported by a Marie Curie research fellowship (European Union
TMR programme) and ARL is supported by the Royal Society. We would
like to thank John Barrow, Eckehard Mielke, Peter Thomas, and Pedro
Viana for helpful discussions and comments. ARL thanks John Webb for
hospitality at UNSW (Sydney) while this paper was being completed.


\nonfrenchspacing


\typeout{IF YOU GET A LaTeX ERROR HERE, READ THE COMMENT AT THE BOTTOM
OF THE LaTeX FILE. OTHERWISE IGNORE THIS! ANDREW}

\begin{table}
\caption[]{The redshift values $z$ for different self-interaction values
$\Lambda $ and initial values of the scalar field $\sigma $. The
addition (max) means the maximal central density up to which stable
boson star states exist for the choice of $\Lambda $.}
\begin{tabular}{ddd}
$\Lambda $ & $\sigma (0)$ & $z$ \\
\tableline
$-$20 & 0.067 & 0.0656984 (max) \\
$-$15 & 0.086 & 0.0869732 (max) \\
$-$10 & 0.122 & 0.129636 (max) \\
$-$5 & 0.195 & 0.237231 (max) \\
0 & 0.025 & 0.0315 \\
0 & 0.271 & 0.456516 (max)  \\
0 & 1.95  & 112.14 \\
5 & 0.253 & 0.565177 (max)  \\
10& 0.225 & 0.610195 (max)  \\
20& 0.184 & 0.642457 (max)  \\
30& 0.158 & 0.652836 (max) \\
50& 0.128 & 0.665739 (max) \\
75& 0.107 & 0.673652 (max) \\
100& 0.0935 & 0.674607 (max) \\
200& 0.0673 & 0.681385 (max) \\
300& 0.0552 & 0.682227 (max) \\
$\Lambda \gg 1 $ & \mbox{$\log \left( \omega /B \right) = 0.287$} & 
0.687385 (max) \\ 
\end{tabular}
\label{table1}
\end{table}

\begin{table}
\caption[]{The maximal rotation velocities at $R$ for different initial
values of $\sigma (0)$ for $\Lambda =10$; cf.~Fig.~\ref{fig5}.}
\begin{tabular}{dddd}
 $\sigma (0)$ & $x_{{\rm max}}$ & $v_{\rm max} \; ({\rm km \, s}^{-1})$ \\
\tableline
0.225 & 5.1  & 113276 \\
0.2   & 5.5  & 108377 \\
0.15  & 6.6  & 95641 \\
0.1   & 8.3  & 78105 \\
0.05  & 11.7 & 54049 \\
0.001 & 81.5 & 7338 \\
0.0001& 257  & 2318 \\
\end{tabular}
\label{table2}
\end{table}


\begin{figure}
\caption[]
{The mass $M$ (---) in units of $(m^2_{\rm Pl}/m)$ and the particle
number $N$ ($--$)in units of $(m^2_{\rm Pl}/m^2)$ as function of the
central value $\sigma (0)$ for different values of $\Lambda := \lambda
m_{{\rm Pl}}^2/4\pi m^2=-5,0,5,10$ \cite{kus,sch}.}
\label{fig1}
\end{figure}

\begin{figure}
\caption[]
{The redshift $z$ depending on the self-interaction constant $\Lambda $
calculated at the first maximum in the ($M, \sigma (0)$) curve
(see Figure \ref{fig1}) which gives the last stable boson star
configuration.}
\label{fig2}
\end{figure}

\begin{figure}
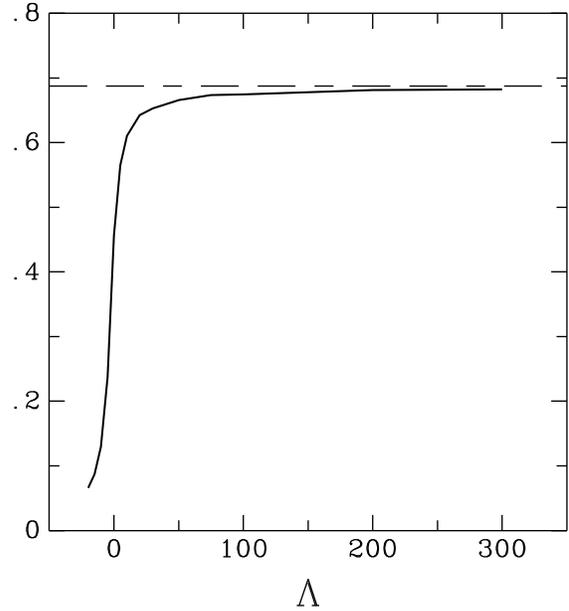

\caption[]
{The redshift function (\ref{zfunc}) for different self-interactions.}
\label{fig3}
\end{figure}

\begin{figure}
\caption[]
{The rotation curves for the 'normal' non-interacting boson star and
with a large value of the constant $\Lambda $. Both curves are taken
for the last stable boson star configuration with the largest mass
value.}
\label{fig4}
\end{figure}

\begin{figure}
\caption[]
{Rotation curves for different initial values $\sigma (0)=0.225, 0.15,
0.05, 0.001$ of $\Lambda =10$; largest initial value has largest
maximal velocity.}
\label{fig5}
\end{figure}


\newpage
\centerline{Figure 1:}
\centerline{\psfig{figure=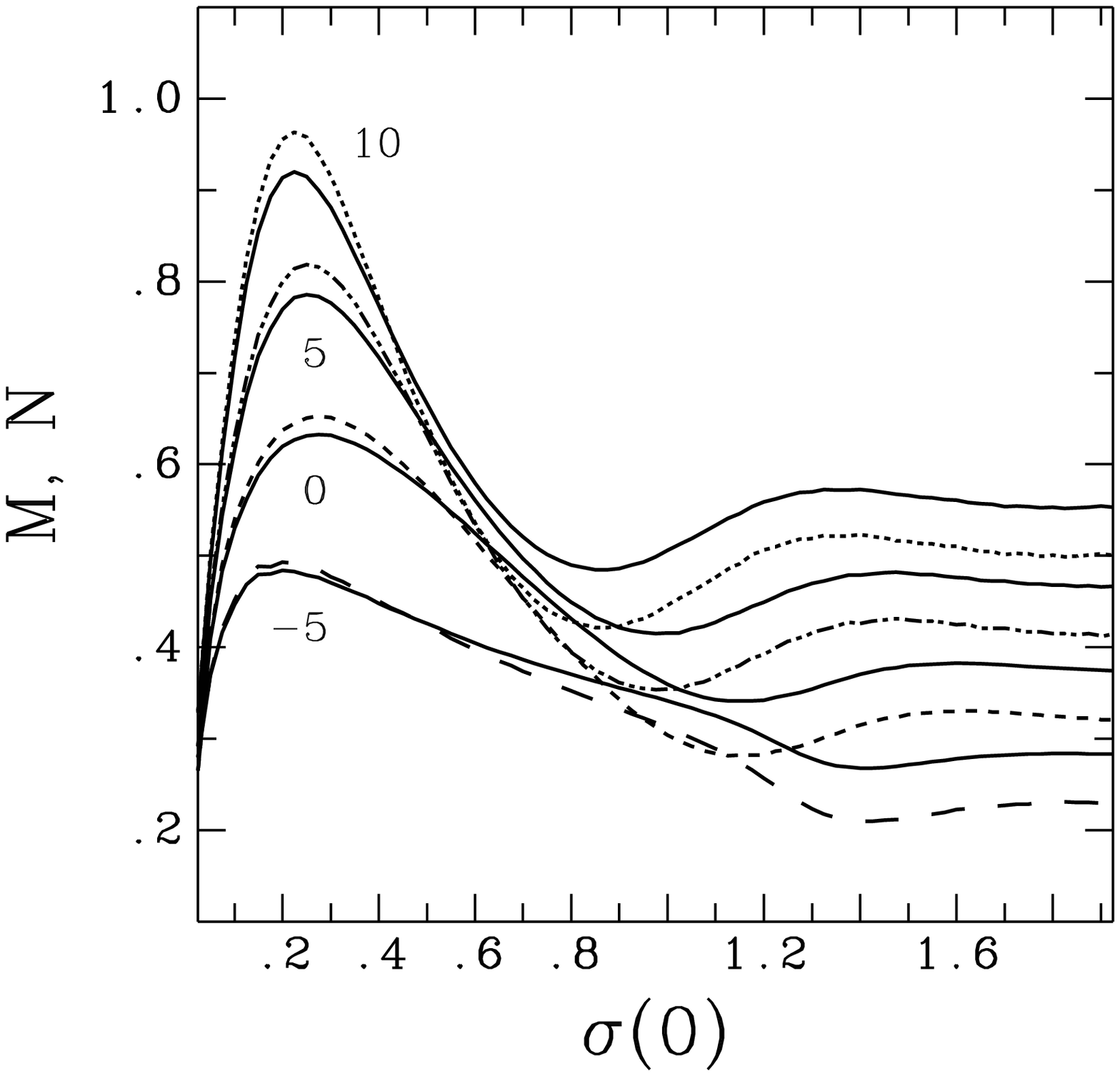,height=8cm} \hskip 2cm}
\vskip 0.5cm

\centerline{Figure 2:}
\centerline{\psfig{figure=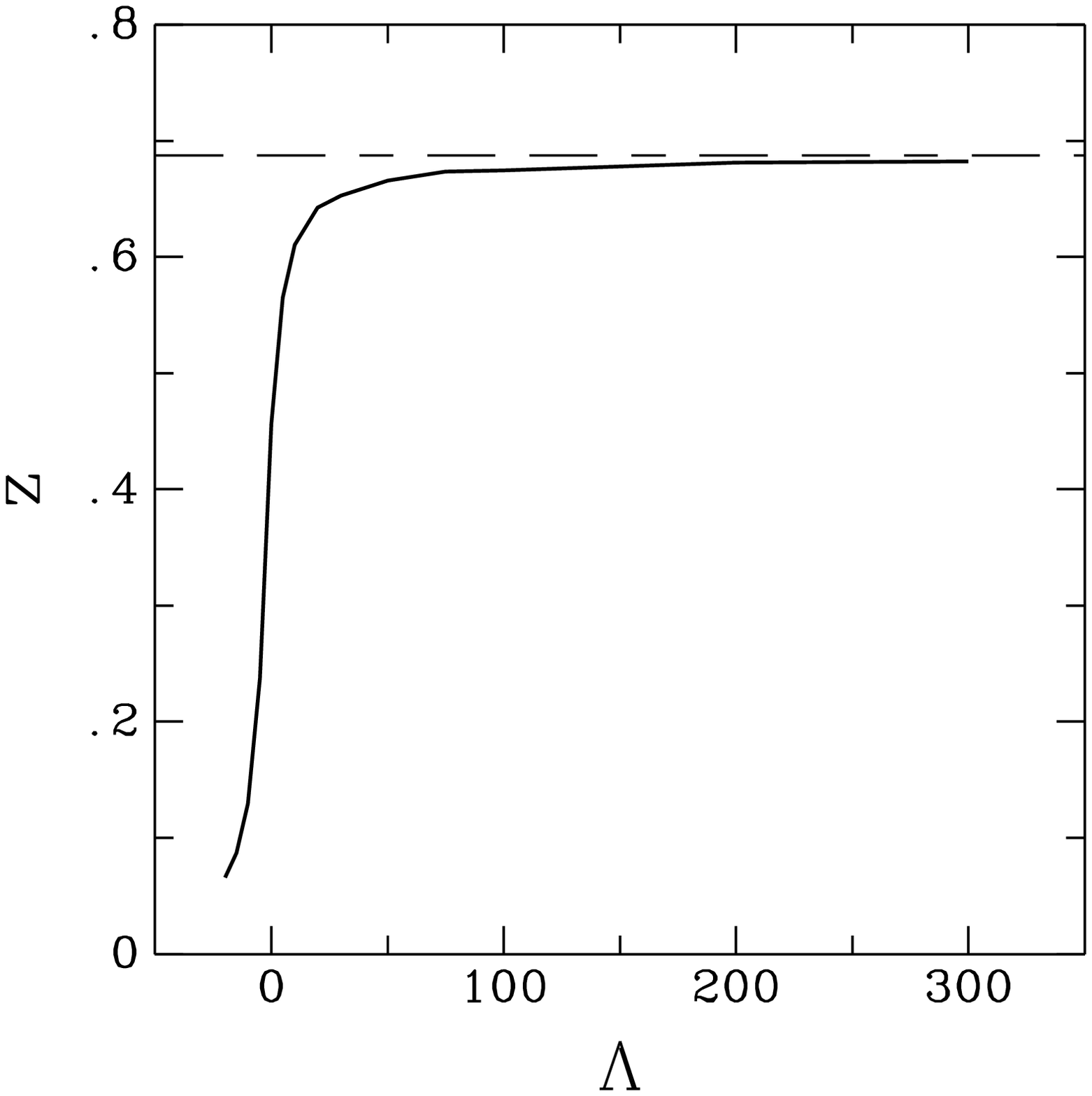,height=8cm} \hskip 2cm}
\vskip 0.5cm

\newpage
\centerline{Figure 3:}
\centerline{\psfig{figure=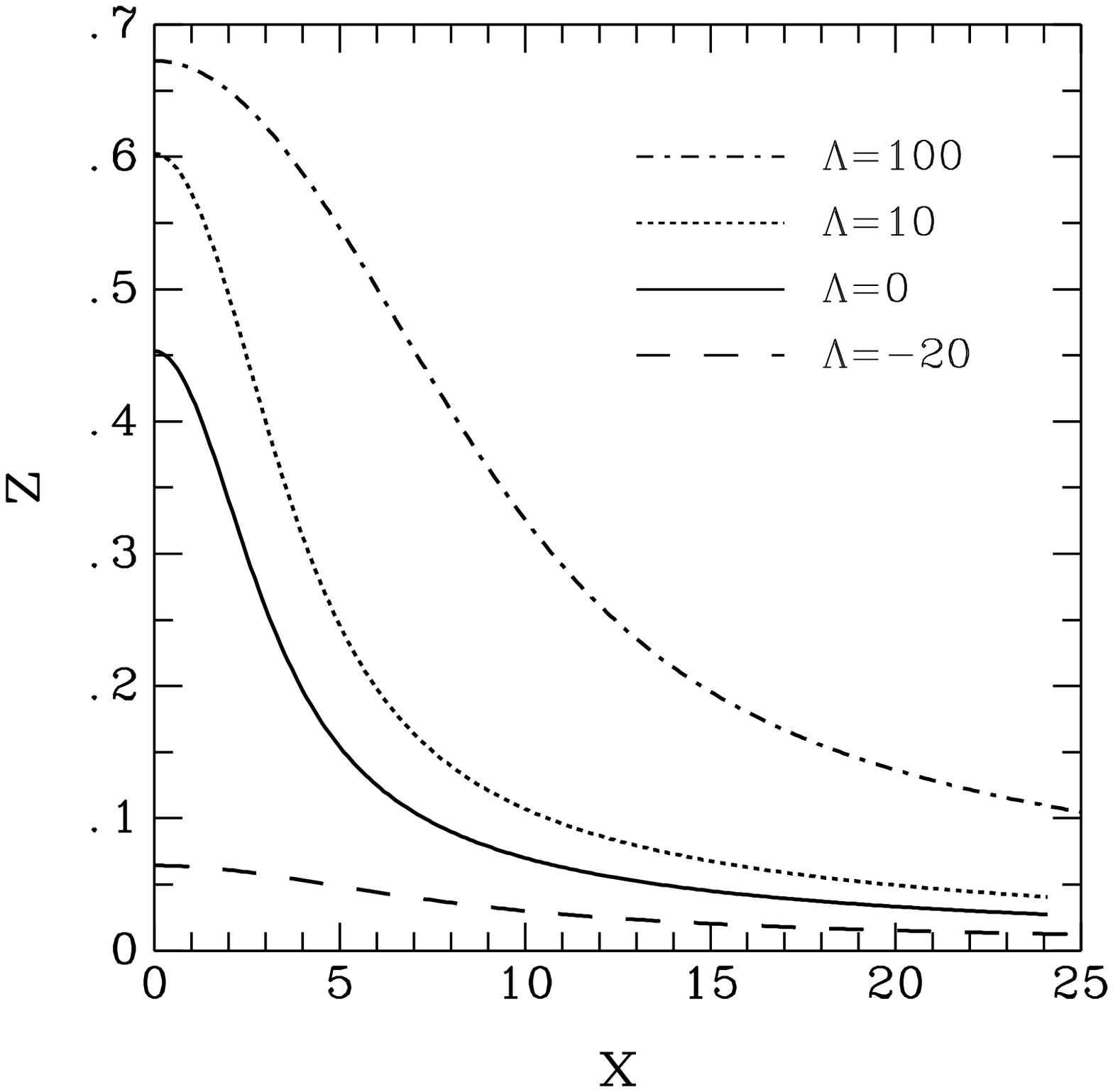,height=8cm} \hskip 2cm}
\vskip 0.5cm

\centerline{Figure 4:}
\centerline{\psfig{figure=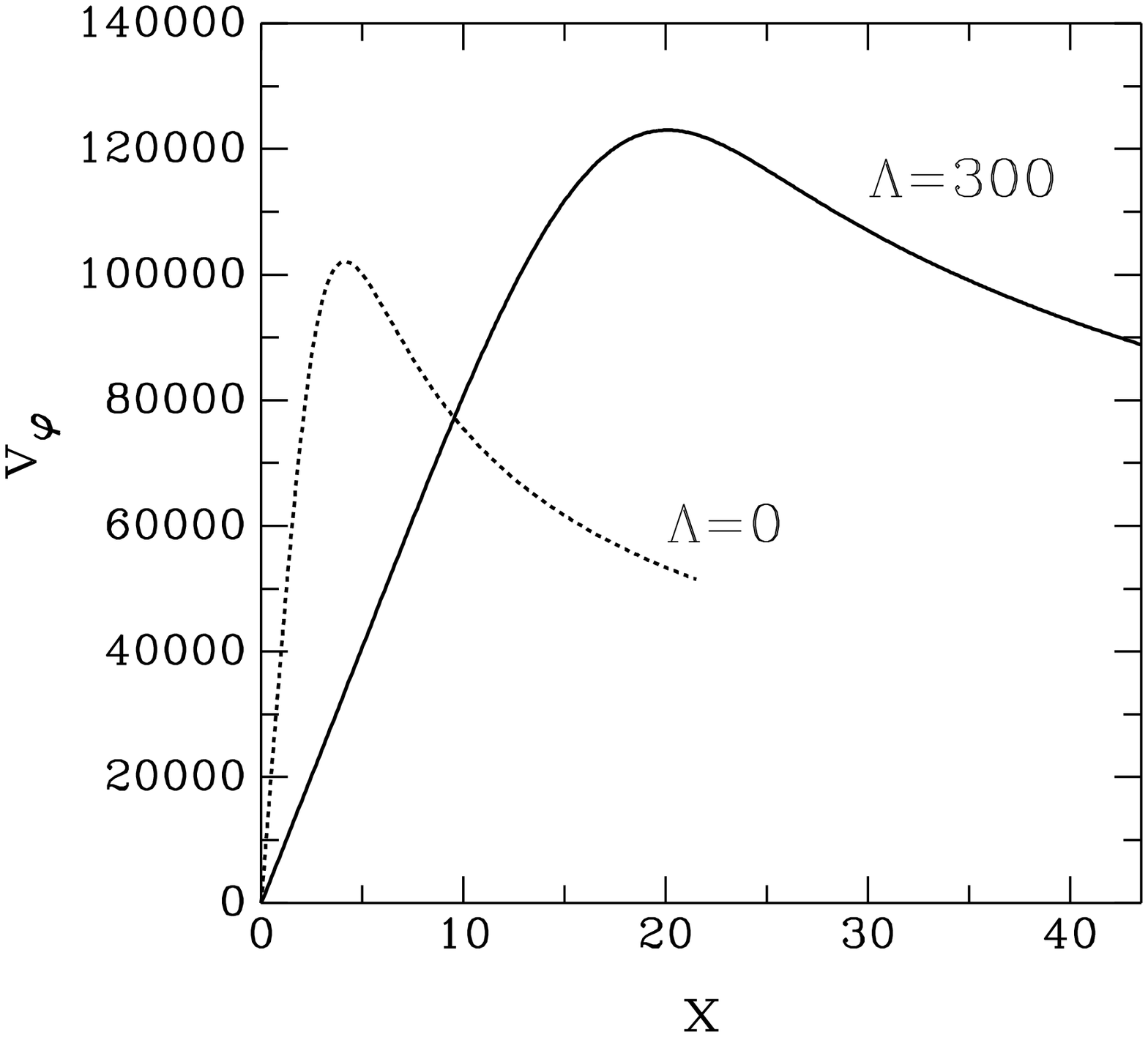,height=8cm} \hskip 2cm}

\newpage
\centerline{Figure 5:}
\centerline{\psfig{figure=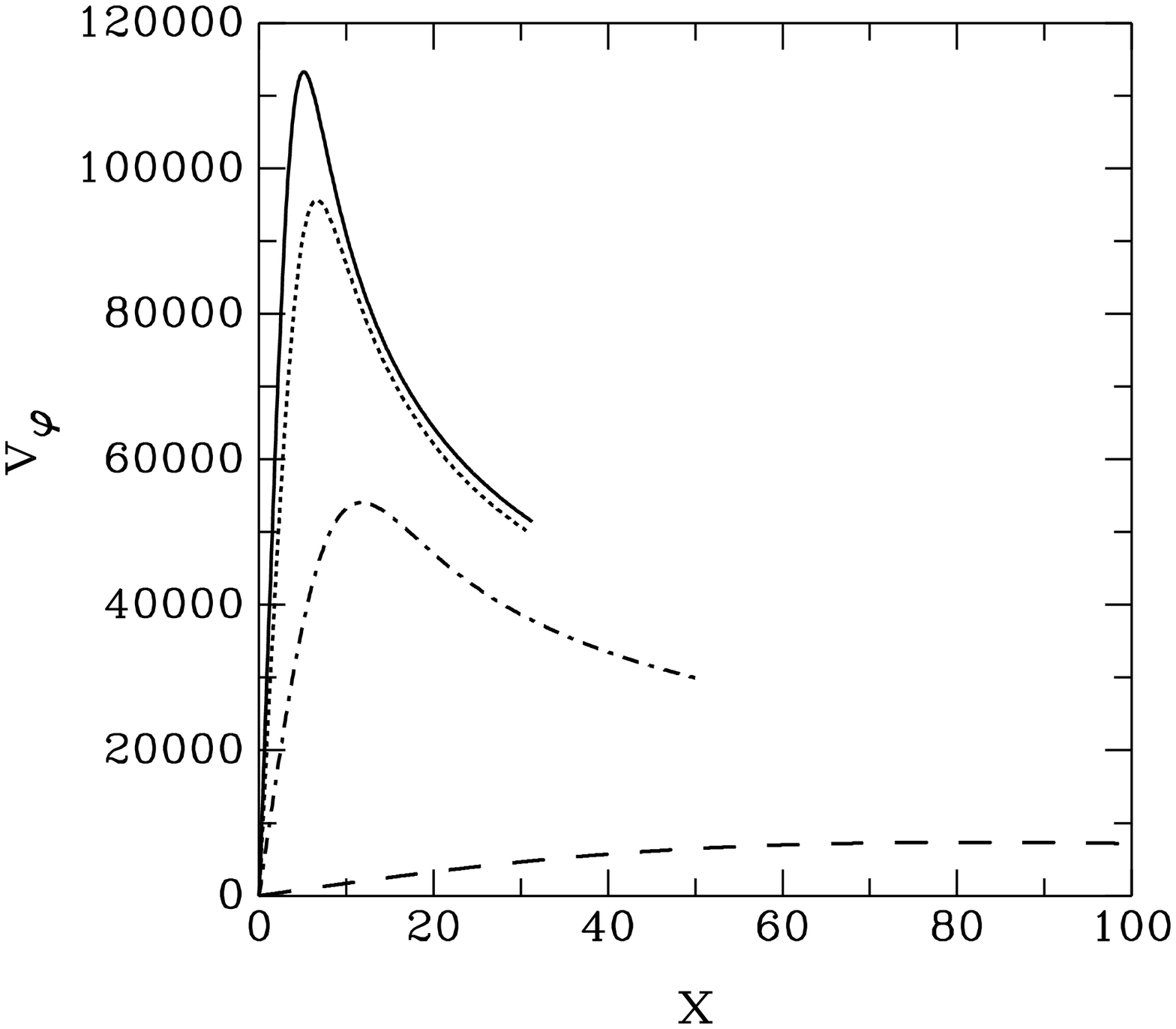,height=8cm} \hskip 2cm}

\end{document}